\documentclass{an}
\usepackage{graphicx}
\usepackage{times}
\Volume{00}              
\Year{2003}              
\Month{00}               
\Pagespan{000}{000}      

\begin{document}
\lhead[\thepage]{B. K\"onig et al.: Flare stars in the TW Hydrae association: 
The HIP~57269 system}
\rhead[Astron. Nachr./AN~{\bf XXX} (2003) X]{\thepage}
\headnote{Astron. Nachr./AN {\bf 32X} (2003) X, XXX--XXX}

\title{Flare stars in the TW Hydrae association: The HIP~57269 system
  \thanks{Data obtained at ESO La Silla and ESO Paranal with programs:
    66.C-0138(B), 67.C-0073(A), 68.C-0009(A), 68.C-0018(A)}}

\author{B. K\"onig
          \inst{1}
          \and
          R. Neuh\"auser\inst{1, 2}
          \and
          E. W. Guenther\inst{3} 
          \and
          V. Hambaryan \inst{4}}
\institute{Max-Planck-Institut f\"ur extraterrestrische Physik,
              Gie\ss enbachstra\ss e 1, D-85748 Garching, Germany
                  \and
              Astrophysikalisches Institut und Universit\"ats-Sternwarte,
              Schillerg\"asschen 2-3, D-07745 Jena, Germany
                  \and
              Th\"uringer Landessternwarte Tautenburg, Sternwarte 5, D-07778
              Tautenburg, Germany
              \and
              Astrophysikalisches Institut Potsdam, An der Sternwarte 16,
              D-14482 Potsdam}
\date{Received {date will be inserted by the editor}; 
accepted {date will be inserted by the editor}} 

\abstract{We discuss a new member candidate of the TW Hydrae association (TWA)
  among the stars of the Gershberg et al.  (\cite{gershberg}) flare star
  catalog. TWA is one of the closest known associations of young stars at
  about 60\,pc. Three supposedly young flare stars are located in the same
  region of the sky as TWA.  One of them (\object{HIP~57269}) shows strong
  Lithium absorption with spectral type K1/K2V and a high level of
  chromospheric and coronal activity. It is located at a distance of
  $48.7\pm6.3$\,pc in common with the five TWA members observed with Hipparcos
  (46.7 to 103.9\,pc). HIP~57268~A has a wide companion~C which also shows
  Lithium absorption at 6707\,\AA~and which has common proper motion with
  \object{HIP~57269}, as well as a close companion resolved visually by
  Tycho. HIP~57269~A\&C lie above the main sequence and are clearly
  pre-main-sequence stars. The UVW-space velocity is more consistent with the
  star system being a Pleiades super cluster member. The two other flare stars
  in the TWA sky region do not show Lithium at all and are, hence, unrelated.
   }

\correspondence{bkoenig@mpe.mpg.de}

\maketitle

\section{Introduction: Flare stars in TW Hydrae} 
The Gershberg et al. (\cite{gershberg}) catalog of UV-Cet type flare star
(FSs) provides a sample of chromospheric and coronal active stars located in
the entire sky.
The existence of luminous young FSs, also in young T associations, indicates a
possible connection between the UV Cet-type and T~Tauri stars.
%
All late-type stars go through a flare stage during their early
evolution. Hence, FSs may well be weak-line or post-TTS or young zero-age
main-sequence stars. The flare rates decline from young pre-main sequence TTS
to somewhat older FSs, but flares on FSs attract more attention because they
do not show other peculiarities (for a study see \cite{guenther} and
\cite{stelzer}).

We placed the stars of the Gershberg catalog in an H-R diagram using their
Hipparcos parallaxes and compared their position to theoretical tracks and
isochrones. In the H-R diagram the stars appear close to the main sequence or
above it. Because of the activity and the position in the H-R diagram, many of
them are likely to be young. To confirm this we take spectra to resolve the
Lithium I line at 6707\,\AA~and compare the line-depth relative to the
continuum to zero-age main-sequence stars of the same spectral type and to
TTSs. Here, we concentrate on flare stars which may be related to the TW
Hydrae association (TWA).  Spectra of all other observed FSs will be published
later.

The star \object{TW Hya} was first thought to be an isolated TTS. Only later
similar stars and member candidates were found by accident and by systematic
searches among infrared and X-ray sources (Hoff et al. \cite{hoff}, Kastner et
al. \cite{kastner}, Jensen et al. \cite{jensen}, Sterzik et
al. \cite{sterzik}, Webb et al. \cite{webb}, Zuckerman et
al. \cite{zuckerman}). More kinematic member candidates can be found in
Makarov \& Fabricius (\cite{makarov}), Torres et al. (\cite{torres}), and
Tachihara et al. (\cite{tachihara}). TWA was found to be a loose group without
nearby cloud material. Today there are 19 known and confirmed member systems
of the association, called TWA 1-19 (Webb et al. \cite{webb}, Zuckerman et
al. \cite{zuckerman}). These stars have common radial velocity and proper
motion and comparable Lithium abundance.  Yet, some discussions are
ongoing, e.g. if TWA-19 is a member of the Scorpio-Centaurus association
(Mamajek et al. \cite{mamajek}). Makarov (2003) has also some restrictions to
TWA-9.

Hipparcos has observed five confirmed member stars of the association. This
leads to a mean distance of $62.2\pm7.8$\,pc with a range of $46.7\pm7.2$\,pc
to $103.9 \pm 17.5$\,pc. The newly proposed member candidate would be the
sixth member with a Hipparcos parallax.

\begin{table*}[ht]
\caption{Basic data of the flare stars located in the TWA sky region} 
\label{sample}
\begin{tabular}{lcccccccc}
\hline
name               & plx                 & V           & SpT.        & $T_{\rm eff}$ & $\mu_{\alpha}$     & $\mu_{\delta}$    & $v \sin i$ & RV \\
                   & [mas]               & [mag]       &             & [K]           & [mas/yr]           & [mas/yr]          &   [km/s]   & [km/s]  \\
\hline
\multicolumn{8}{l}{Flare stars located in the TWA-region}\\
\hline
\object{HIP 57269} & 20.55$\pm$2.38$^1$  & 9.34$^{11}$ & K1/K2V$^9$  &
$4990\pm100^{12}$  & $-137.21\pm1.90^1$ & $-47.71\pm1.37^1$ & $20^7$ & $16^8$,
$19.0\pm3.0^{13}$\\
\object{HIP 56244} & 95.46$\pm$2.59$^1$  & 11.5$^4$    & M3.5e $^2$  & $3420\pm100^{12}$  & $-715.75\pm1.38^1$ & $170.75\pm1.38^1$ &  &  $6.0^6$\\
\object{GJ 3780}   &  $62.5\pm14.4^3$    & 12.9$^4$    & M3.5$^2$    & $3420\pm100^{12}$  & $-556^5$           & $-249^5$          &  &  $-0.3^6$\\
\hline
TWA range$^{4,10}$ & 9.62 to 21.43       & 8 to 14     & A0 to M9    &               & -122.2 to -30.0    & -38.0 to +16.3    & 5 to 58 & 5.73 to 14.0 \\
\hline
\end{tabular}

\vspace{0.1cm}
1: Hipparcos catalog
2: Gershberg et al. (1999)
3: Hawley et al. (1997)
4: SIMBAD
5: Bakos et al. (\cite{Bakos})
6: Gliese \& Jahreiss (\cite{Gliese}): Catalog of nearby stars (CNS4)
7: Pallavicini et al. (1992)
8: Anders et al. (1991)
9: Cutispoto (1998)
10: Torres et al. (2003)
11: Fabricius \& Makarov (\cite{fabricius2000})
12: Adapted from Kenyon \& Hartmann (\cite{kenyon})
13: this work, measured from two DFOSC spectra
\end{table*}

\section{Spectroscopy in the optical regime}
The optical spectroscopy was performed using the Danish 1.54\,m telescope at La
Silla, ESO equipped with the focal reductor instrument DFOSC (Danish Faint
Object Spectrograph and Camera) in short slit \'echelle mode. The spectral 
resolution in the red wavelength range is about 5000 
($\lambda/\Delta \lambda$). This resolution
is sufficient to distinguish Lithium I at 6707\,\AA~from Calcium at
6718\,\AA, but not from the nearby much weaker Iron lines at 6706\,\AA. The
total wavelength range covered is from 5850\,\AA~to 8500\,\AA, i.e. including
H${\alpha}$.
  
Data reduction was performed using the \'echelle package of IRAF. We have
subtracted the mean bias created with dedicated bias images and a flat-field
correction was applied, using a mean dome-flat. We note that the flat-field
lamp does not have a Lithium I line at 6707~\AA. In each spectra we subtracted
the sky by extracting 10-13 pixel to the left and right of the stellar
spectrum.

The sample of stars were taken from Gershberg et al. (\cite{gershberg}) 
and listed in Table~\ref{sample}, namely the stars located in the region of
$\alpha = 10$\,h to 13\,h and $\delta = -24^\circ~30^m$~to $-50^\circ$.

\begin{figure}[h]
\resizebox{\hsize}{!}
{\includegraphics[angle=90]{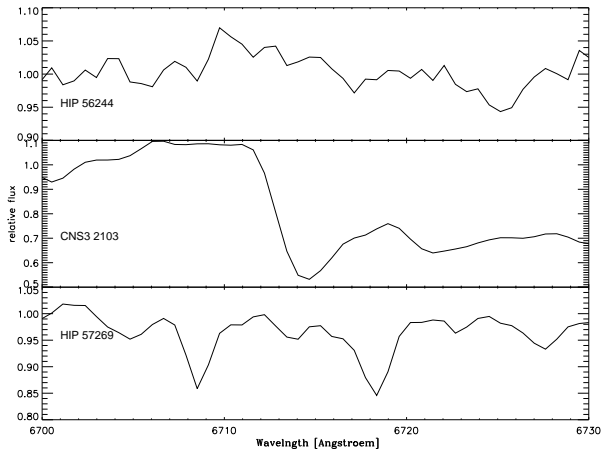}}
\caption{A section of the spectrum of \object{HIP 57269 A\&B} showing Lithium
  at 6708\,\AA~and Calcium at 6718\,\AA~and \object{HIP 66244} and
  \object{GJ~3780} without Lithium.}
\label{fig2}
\end{figure}
\begin{figure}[h]
\resizebox{\hsize}{!}
{\includegraphics[angle=90, width=0.5\textwidth]{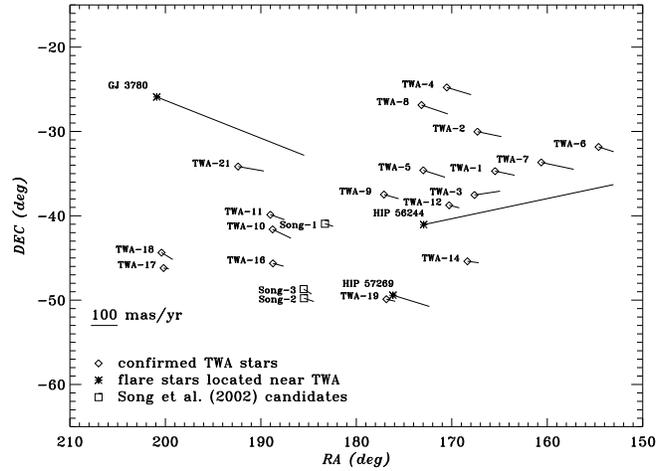}}
\caption{Proper motion of the known confirmed TWA members and the flare stars
  discussed in this paper: \object{HIP~57269}, \object{GJ~3780} and
  \object{HIP 56244}. Proper motions were taken from Torres et al. (2003) for
  the TWA members and the Song et al. (2002) member candidates, for the flare
  stars from the Hipparcos catalog or from the Simbad database. The labels
  song-1 correspond to TYC~7760-0835-1, song-2 to TYC~8238-1462-1, and song-3
  to TYC-8234-2856-1.}
\label{fig1}
\end{figure}
\begin{figure*}[ht]
\resizebox{\hsize}{!}
{\includegraphics[angle=270, width=\textwidth]{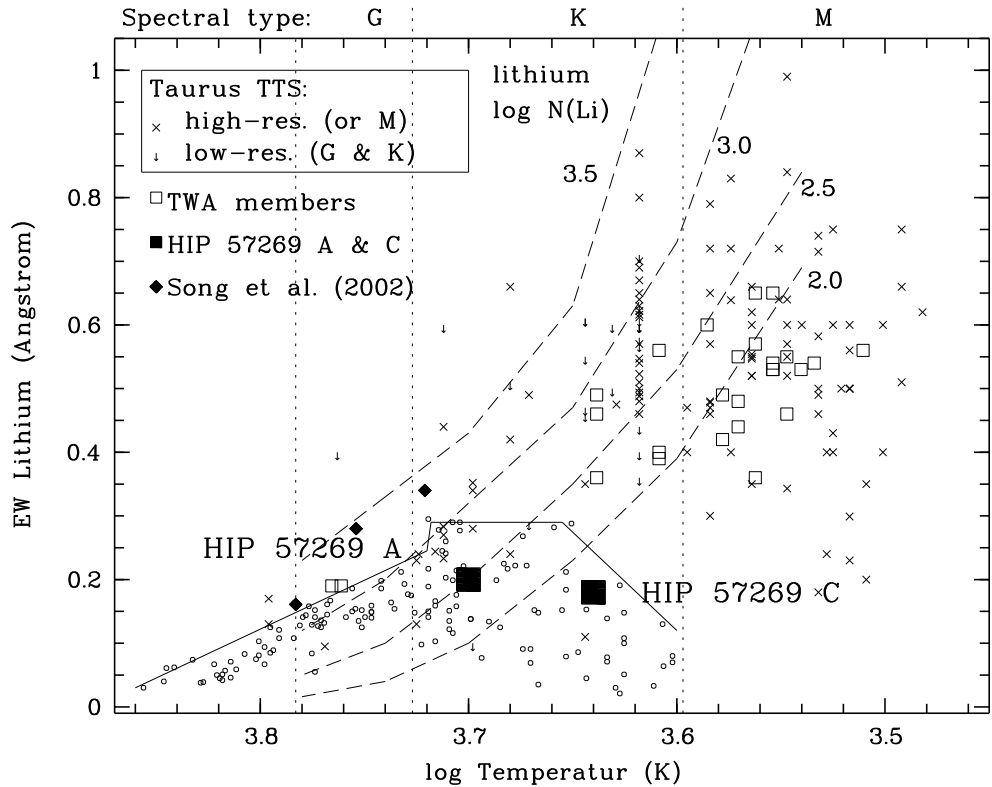}}
\caption{The Lithium EW versus temperature of the confirmed TWA member stars
  (open boxes) and \object{HIP~57269 A} and its companion C (filled box). The
  three member candidates from Song et al. (2002) are also shown (big dots)
  with temperatures adopted from Mamajek et al. (2002) for two stars and for
  one estimated from Kenyon \& Hartmann (1995). Underlayed are TTSs in Taurus
  (crosses and arrows(upper limits)) and Pleiades stars (open dots) for
  reference. Figure adapted for the TTS and Pleiades stars from Neuh\"auser
  (\cite{neuhaeuser1997}). The temperatures for the TWA members and member
  candidates were derived from published spectral types using the Kenyon \&
  Hartmann (\cite{kenyon}) temperature scale. Lithium iso-abundance lines are
  taken from Pavlenko \& Magazz\`u (\cite{pavlenko}).}
\label{fig:LivsTemp}
\end{figure*}
\begin{figure*}[ht]
\resizebox{\hsize}{!}
{\includegraphics[angle=90, width=\textwidth]{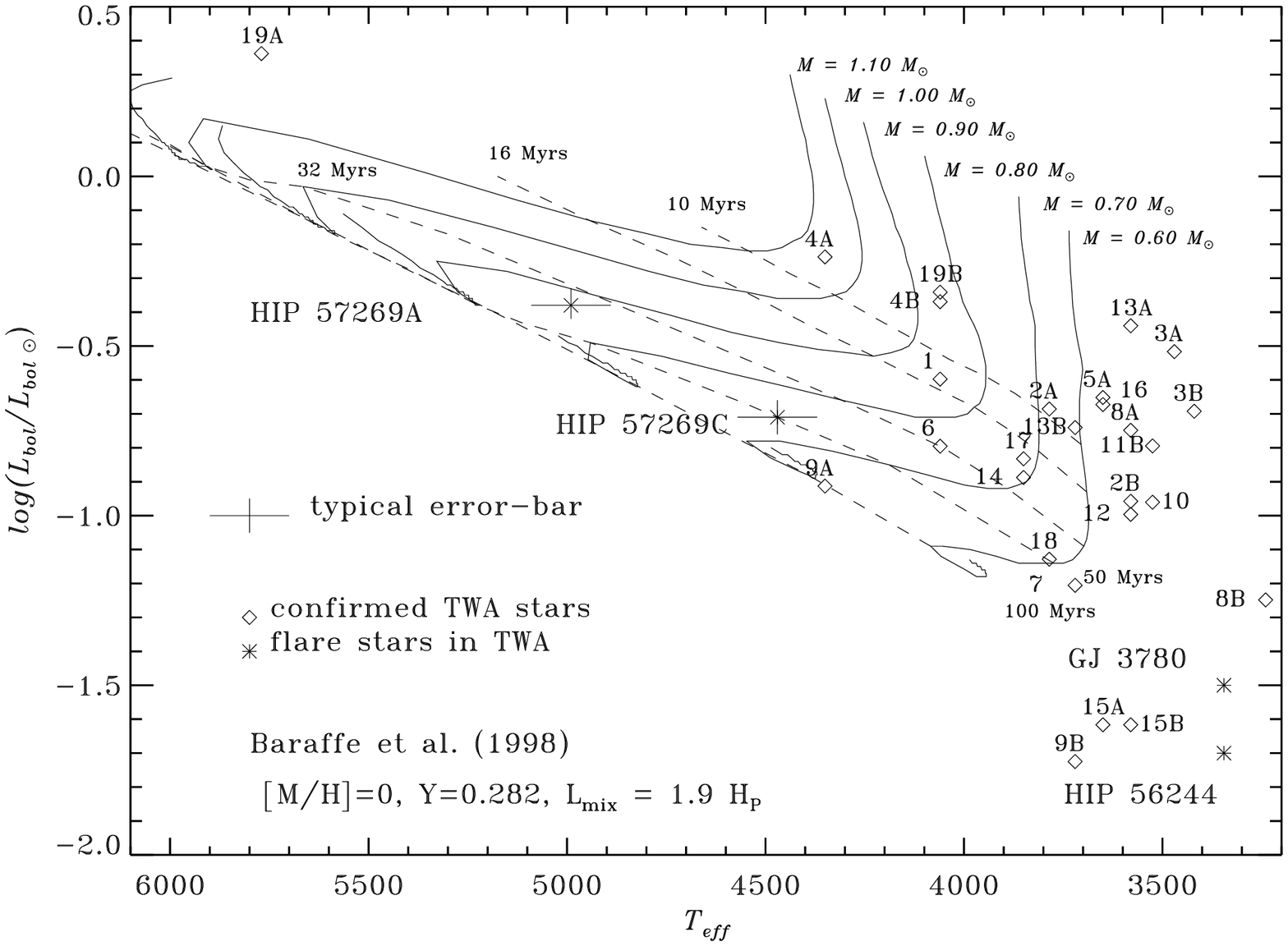}}
\caption{The three flare stars and HIP~57269~C (big asterix) together with the
  confirmed members of TWA (TWA-1 - 19) plotted in an H-R diagram using
  distances from Tab.~\ref{sample}. The luminosity for HIP~57269~A was
  computed using Tycho $V_T$- and $B_T$-magnitudes for the component A
  only. The V-magnitude for component C was taken from SIMBAD. For a
  comparison are overlayed the theoretical tracks and isochrones of Baraffe et
  al. (\cite{baraffe}). It can be seen that both HIP~57269~A \& C lie above
  the main sequence and on the same isochrone taking into account the errors
  in photometry and conversion to temperature from their spectral type. The
  temperature for all stars in the diagram were derived using Kenyon \&
  Hartmann (\cite{kenyon}). The photometry was taken from various sources (the
  references can be found in Torres et al.~\cite{torres2003})
  and the distance to the individual objects was taken from Hipparcos or from
  Frink (\cite{frink}) when possible, for the other stars we use a mean
  distance of 61.5\,pc. GJ~3780 and HIP~56244 are unrelated foreground
  stars. TWA-9~B and TWA-15~A\&B are also located on the lower right below the
  main-sequence, most certainly because the assumed mean distance of 61.5\,pc
  to TWA may be different to the true distance of these stars.}
\label{fig:DAM}
\end{figure*}
The two stars {\object{HIP~56244}} and {\object{GJ~3780}} do not show Lithium
(see Fig~\ref{fig2}). HIP~56244 is classified as a flare star by
\cite{pettersen} and GJ~3780 by \cite{eggen1987}. Their proper motion
(Fig.~\ref{fig1}) is not consistent 
with the mean proper motion of the TWA member stars (see Fig.~\ref{fig1}:
$\mu_{\alpha}=-80.9\pm5$\,mas/yr, $\mu_{\delta}=-26.4\pm5$\,mas/yr, Webb et
al. \cite{webb}). {\object{HIP~56244}} is a foreground object at a distance of
$10.48\pm0.28$\,pc. Hawley et al. (\cite{Hawley1997}) give a spectroscopic
distance to {\object{GJ~3780}} of $16.0\pm4.8$\,pc, which also leads us to the
conclusion that it is a foreground star.
\begin{table}[h]
\caption{Spectroscopic data of HIP~57269}
\label{tab:li}
\begin{tabular}{lccccc}
\hline
W$_{\lambda}$(Li) & W$_{\lambda}$(Ca) & log N(Li) & $v \sin{i}$ &  rad. vel. &ref. \\
 $[$m\AA]   &  [m\AA]          &           & [km/s]    & [km/s]    &      \\    
\hline
196$\pm$5               & 200$\pm$3  & 2.7       & 20     &              &  P   \\
                        &            & 2.2       & 20     &              &  R   \\
205                     &            & 2.5       & 23     & 15.9         &  A   \\
196                     &            &           & 20     &              &  S   \\     
200$\pm20$              & 270$\pm20$ & 2.5       &        & $19.0\pm3.0$ &  K   \\
\hline
\end{tabular}

P: Pallavicini et al. (\cite{pallavicini}), R: Randich et al. (\cite{randich}),
A: Anders et al. (\cite{anders}), S: Song et al. (\cite{song}), K: this paper
\end{table}
\cite{udalski} first classified HIP~57269 as a BY~Dra type variable.
\object{HIP~57269 A}, a star with spectral type K1/K2V, shows Lithium
absorption (see Fig.~\ref{fig2}). The spectral type was published by Cutispoto
(\cite{cutispoto}) and references therein after a thorough study and was
confirmed by our DFOSC spectrum. The equivalent width (EW) of the
6707\,\AA~line is $0.20\pm0.01$\,\AA~while Calcium at 6718\,\AA~has
$0.27\pm0.01$\,\AA. This is very similar to the Lithium EW of TWA-19 with
spectral type G3-5, with 0.19\,\AA~EW (Webb et al. \cite{webb}, Sterzik et
al. \cite{sterzik} and Zuckerman et al. \cite{zuckerman}), see
Fig.~\ref{fig:LivsTemp}. This Lithium EW and spectral type leads to a Lithium
abundance of $\log({\rm N}(Li)) = 2.5$ for \object{HIP~57269} following
Pavlenko \& Magazz\`u (\cite{pavlenko}) for dwarfs. This $W_\lambda(Li)$~is
near the upper envelope of $W_\lambda(Li)$~of the Pleiades at the spectral
type early K and similar as in G- and early K-type TTS in Taurus (see
Fig~\ref{fig:LivsTemp}).
 
Pallavicini et al. (\cite{pallavicini}) and Randich et al. (\cite{randich})
observed \object{HIP~57269} with high resolution ($\Delta\lambda/\lambda
=50,000$) and analyzed it in their sample of chromospheric active stars and RS
CVn stars. Randich et al. (\cite{randich}) conclude that it is a young
chromospheric active star and not an old RS~CVn star, see Tab.~\ref{tab:li}.

Anders et al. (\cite{anders}) have also analyzed high resolution spectra of
this star deriving high Lithium abundances. They also take into account the
proper motion, the space motion and a distance of 42\,pc, they conclude that
the star is a Pleiades super cluster member. We note that before 1997, when
Pallavicini et al. (\cite{pallavicini}), Randich et al. (\cite{randich}), and
Anders et al. (\cite{anders}) studied this star, TWA was not yet recognized as
a young cluster and the proper motion of the Pleiades super cluster is quite
similar to the proper motion of TWA.

\section{Kinematics and multiplicity}
The proper motion of \object{HIP~57269} ($-137.21\pm1.0$\,mas/yr,
$-47.71\pm1.37$\,mas/yr) could be consistent with the proper motion (see
Fig.~\ref{fig1}) of the other stars in TWA, so that Makarov \& Fabricius
(\cite{makarov}) list it as a kinematic member candidate. 
The radial velocity measured by us of $19.0\pm3.0$\,km/s and the space
motion leads to a space velocity of $U=18.5$\,km/s, $V=-28.4$\,km/s and
$W=-15.2$\,km/s which is quite different from the average space velocity of
TWA ($U=10.8$\,km/s, $V=17.7$\,km/s, $W=-5.6$\,km/s, Torres et
al. \cite{torres2002}) and it is more consistent with the star being a
Pleiades super cluster member (Chereul et al.\cite{cheureul}).

HIP~57269 (component A: V=9.34\,mag) is a common proper motion triple
star with the secondary (component B) being a V=10.48\,mag companion
at a position angle of $306^{\circ}$~and 0.430" separation (Fabricius
\& Makarov \cite{fabricius2000}) and the third star (component C) with
V=13.5\,mag at a position angle of $349^{\circ}$~and 9.6" separation
(Hipparcos catalog) (see also Fig.~\ref{fig4}). The component C also
shows Lithium absorption at 6707\,\AA~(Fig.~\ref{fig3}). Multiplicity
is typical for young stars.

\section{H-R diagram}
We use the Tycho $V_T$- and $B_T$-magnitudes for component A
(Fabricius \& Makarov~\cite{fabricius2000}) only to convert to
luminosity, and to place it into the H-R diagram at a temperature of
$4990 \pm 100$\,K at a spectral type of K1/2V.  In the H-R diagram
(Fig.~\ref{fig:DAM}) HIP~57269~A lies above the main sequence, with an
age of $40\pm10$\,Myrs and a mass of $0.88\pm0.05\,{\rm
  M}_\odot$~compared to tracks and isochrones of Baraffe et al.
(\cite{baraffe}). This age is comparable to the age of TWA-19A.

We also include the wide companion HIP~57269~C in the diagram.
Assuming that the star is at the same distance as the component A, the
star has an age of $50\pm10$\,Myrs, consistent with A and a mass of
$0.75\pm0.05\,{\rm M}_\odot$. This star is clearly younger than the
TWA-9A and it lies on the same isochrone as TWA-7 and TWA-18.

Song et al. (\cite{song}) argue that \object{HIP 57269} not a member
of TWA because the Li EW is more consistent with a 30\,Myrs star
rather than a 10\,Myrs, and that it lies on the main sequence on a
specific set of the Siess et al. (\cite{siess}) models.

Fabricius \& Makarov~(\cite{fabricius2000}) give Tycho $B_T$~and
$V_T$~for the components HIP~57269 A \& B separately. Transforming
those to a Johnson V-magnitude and calculating the absolute magnitude,
HIP~57269 A appears above the main-sequence on the 30\,Myrs isochrone
using the Siess et al. (\cite{siess}) model shown in the paper of Song
et al. (\cite{song}). Given the error-bars, the star may be consistent
with an age of 10\,Myrs in the Siess et al. (\cite{siess}) tracks.

\section{X-ray emission}

The region of TW Hydrae was observed during the ROSAT all-sky survey (RASS)
using the Position-Sensitive Proportional Counter (PSPC) in scanning mode. A
total of 127\,sec have been taken, making it possible to detect
\object{HIP~57269} clearly with a maximum likelihood (ML) of 268.7. The source
has a count-rate of $0.75\pm0.08$\,cts/sec. The count-rate range for single or
unresolved TWA members is 0.11\,cts/sec to 0.66\,cts/sec (Stelzer \&
Neuh\"auser \cite{stelzer2000}). The triple HIP~57269 appears to be the X-ray
brightest TWA member (candidate), not surprising, because it is an unresolved
multiple in the RASS, it has a relatively close distance, it shows activity
(as flare star), and it has one of the earliest spectral types in TWA. A high
level of X-ray activity is a strong youth signature.

The PSPC provides some spectral
information, allowing to calculate hardness ratios:
\begin{equation}
HR1 = \frac{cr_{H1}+cr_{H2}-cr_S}{cr_{H1}+cr_{H2}+cr_S}
\end{equation}
\begin{equation}
HR2 = \frac{cr_{H2}-cr_{H1}}{cr_{H1}+cr_{H2}}
\end{equation}
where $cr_{S}, cr_{H1}, cr_{H2}$~denote the count-rates in the three
ROSAT-PSPC energy bands soft (0.1 - 0.4\,keV), hard 1 (0.4 - 0.9\,keV) and
hard 2 (0.9 - 2.1\,keV) respectively. See Tab.~\ref{xray} for the ROSAT
data. These hardness ratios for \object{HIP~57269} are typical for TWA (HR1:
-0.31 to 0.58 and HR2: -0.29 to 0.53, Stelzer \& Neuh\"auser
\cite{stelzer2000}).

The RASS observations of field rs932622 do not show flare activity for
HIP~57269. Follow-up XMM spectroscopy and variability monitoring as well as
high angular resolution X-ray imaging by Chandra would be interesting.

\begin{table}[h]
\caption{X-ray data of the FSs located in the sky region of TWA.}
\label{xray}
\scriptsize
\begin{tabular}{lcccc}
\hline
name               & RASS    & ML & HR1 & HR2 \\
                   & cts/s   &    &     &     \\
\hline
\multicolumn{5}{l}{RASS-fields rs932622, rs932233 and rs932425}\\
\hline
\object{HIP 57269} & $0.75\pm0.08$ & 268.7 & $-0.27\pm0.10$ & $0.04\pm0.18$ \\
\object{HIP 56244} & $0.79\pm0.06$ & 600.0 & $-0.35\pm0.07$ & $0.27\pm0.27$ \\ 
\object{GJ 3780}   & $0.15\pm0.02$ &  71.6 & $-0.28\pm0.16$ & $0.05\pm0.27$ \\
\hline
\end{tabular}
\end{table}
 
\section{Near-infrared imaging of HIP 57269}
\begin{figure}
\resizebox{\hsize}{!}
{\includegraphics[angle=0,width=8.5cm, height=8cm]{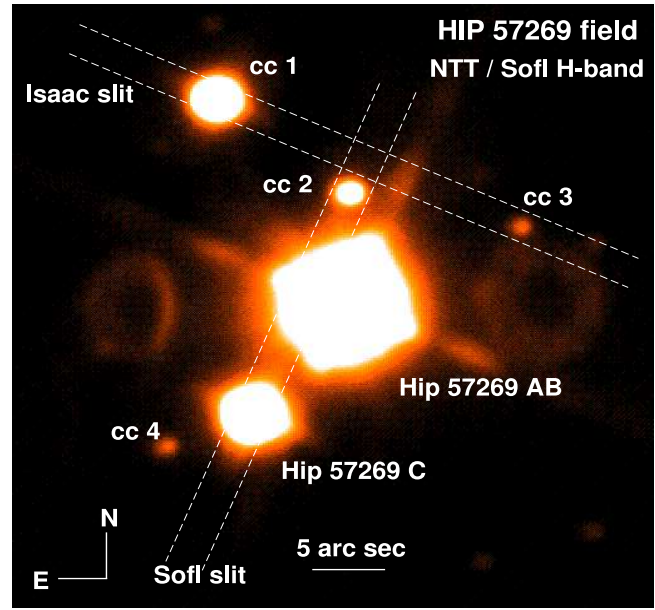}}
\caption{HIP 57269 A with companion C and a few more companion candidates
labeled cc 1 to cc 4 detected in our SofI H-band image; superimposed are the
slit orientation for follow-up spectroscopy with SofI in the same run and
ISAAC later on. Slit widths are not to scale.}
\label{fig4}
\end{figure}

We imaged \object{HIP 57269} in the H-band with the Son of Isaac
(SofI\footnote{see www.ls.eso.org/lasilla/Telescopes/NEWNTT/}) at the
3.5\,m New Technology Telescope (NTT) of the European Southern
Observatory (ESO) on La Silla, Chile, on 2001 Dec 8 from 08:38\,h to
08:58\,h UT with 500 times 1.2\,s integrations. The SofI detector is
an Hawaii HgCdTe $1024 \times 1024$ array with $18.5\,\mu$m pixel
size. We used the small SofI field with its best pixel scale for
better angular resolution and determined the pixel scale by comparing
the separations between several stars on other images taken in the
same night with 2MASS images of the same fields to be $0.150 \pm
0.002^{\prime \prime}$ per pixel. Darks, flats, and standards were
observed in the same nights with the same set-up and data reduction
was done with {\em eclipse}\footnote{see
  www.eso.org/projects/aot/eclipse/} version 3.8, a C-based software
library. While {\em eclipse} is made for VLT data reduction, like e.g.
the Infrared Imaging And Array Camera (ISAAC), and not guaranteed to
work for SofI data, it also does work for SofI imaging data reduction
(dark, flat, shift+add); after all, SofI is the Son of Isaac. See
Fig.~\ref{fig4} for the final co-added image of \object{HIP~57269} and
its surroundings. The FWHM in the final image is 0.9".

The bright object 8.5" SE of \object{HIP 57269 AB} is the known wide
companion HIP~57269~C. We label the additional companion candidates cc~1 to
cc~4 (cc for companion candidate) considering only those candidates within
(somewhat arbitrary) 1000 AU, i.e. 16.6".
The faint HST standard stars S361-D and S754-C (Persson et al. \cite{persson})
were used to obtain the H-band magnitudes of \object{HIP~57269~AB} and C
as well as its companion candidates, see Tab.~\ref{tab:Hmag} ($\pm 0.2$ mag).
Whether the companion candidates are truly bound companions to \object{HIP
57269} can be decided on the bases of spectroscopy and/or proper motion
follow-up observations. 2MASS data of this field are not yet available.
\begin{table}[h]
\caption{HIP 57269 and companion candidates}
\begin{tabular}{lrrlrc}
\hline
Name & \multicolumn{3}{c}{Separation ["]} & H & Spec \\
     & $\Delta \alpha$ & $\Delta \delta$ &  & [mag] & type \\ \hline
HIP 57269 A    & \multicolumn{3}{c}{primary} & 7.1$^\star$ & K1/2V \\ \hline
HIP 57269 C    &  5.11 & -6.74 & SE &  9.6 & K4-6 \\
HIP 57269/cc 1 &  7.18 & 12.48 & NE & 10.7 & K5-7 \\
HIP 57269/cc 2 & -0.43 &  6.74 & NW & 12.8 & G-M \\
HIP 57269/cc 3 &-10.22 & -4.65 & NW & 15.3 & mid-K \\
HIP 57269/cc 4 & 10.15 & -8.68 & SE & 15.2 & ? \\ \hline
\end{tabular}

$\star$: Combined H-band magnitude of A \& B 
\label{tab:Hmag}
\end{table}

\section{Spectroscopic follow-up of companion candidates}
To verify or reject the companion candidates as real companions or
unrelated background objects, we have taken follow-up spectra, both in
the optical and in the infrared (for cc~1 through cc~3).

Optical spectra have been obtained with DFOSC at the 1.54\,m Danish telescope
located at ESO La Silla on 2002 January 24th for both \object{HIP~57269~C} and
\object{HIP 57269/cc~1} to determine their spectral type and to check for
Lithium absorption, a youth indicator, which should be present, if the objects
were real companions, i.e. as young as the primary.

\begin{figure}
\resizebox{\hsize}{!}
{\includegraphics[angle=90]{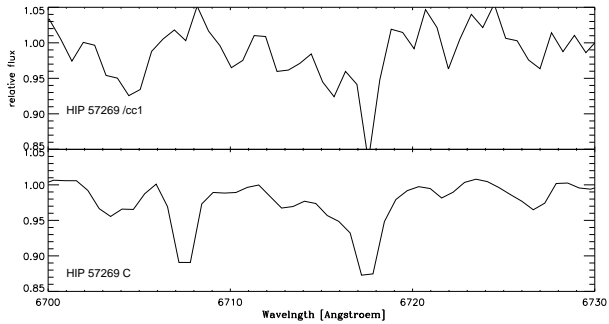}}
\vspace{-2cm}
\caption{A section of the optical spectrum taken with DFOSC at La Silla of
  \object{HIP~57269~C} and cc~1. Lithium is detected in the
  \object{HIP~57269~C}.}
\label{fig3}
\end{figure}

Figure~\ref{fig3} shows part of the spectrum taken with DFOSC and the same
setup as for HIP~57269~A of the companion at the separation of 8.46". In the
spectrum Lithium and Calcium are again clearly seen. With a Lithium EW of
$0.18\pm0.20$\,\AA~at a spectral type of K5V quite similar to the component A.
We have also taken an optical spectrum with DFOSC of cc~1. The spectrum
reveals that the object is a star with spectral type K5-K7 not showing Lithium
at all. This leads to the conclusion, that cc1 is not a member of the
\object{HIP~57269} system and also not of TWA.
\begin{figure}
\resizebox{\hsize}{!}
{\includegraphics[angle=270,width=0.5\textwidth]{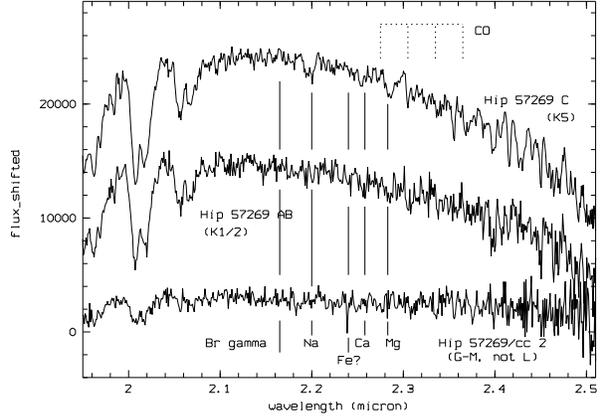}}
\caption{Infrared spectra taken with SofI. For alignment of the slit see
  Fig.~\ref{fig4}. We determine the spectral type of component C to K5$\pm 1$,
  and of cc~2 to G-M, not L as the H-band magnitude suggests, if cc~2 would be
  a true companion.}
\label{fig:specSOFI}
\end{figure}

Infrared spectra for \object{HIP 57269 C} and cc~2 on the slit (together with
HIP~57269~A itself off the slit) have been obtained with SofI in the night 9
Dec 2001 between 08:35\,h and 08:56\,h UT. We took 40 spectra with 30\,sec
exposure each through a $1^{\prime \prime}$ slit with a red grism including
both the H- and K-band (1.53 to $2.52\,\mu$m) with a resolution of $R \simeq
1000$.  Data reduction was done in the usual way using IRAF: Dark subtraction,
normalization, flat fielding, sky subtraction, wavelength calibration, and
co-adding the spectra.  The spectra were not flux-calibrated. The final K-band
spectra of \object{HIP~57269~A}, C, and cc~2 are shown in
Fig.~\ref{fig:specSOFI}: HIP~57269~A is known to be a K1/K2-type dwarf star
(see Fig.~\ref{fig3}), and the spectral type of \object{HIP~57269~C} was just
determined to be K5$\pm 1$ by us (Fig.~\ref{fig3}).

Our IR spectra are consistent with those spectral types, we can see the
typical Na, Mg, and Ca lines as well as weak CO molecular bands, but no Br
$\gamma$ lines which would be typical for earlier types.  Unfortunately, the
spectrum of cc~2 is very noisy, so that it is hard to determine the spectral
type. Na and Br $\gamma$ are very weak or not present at all, CO bands are
also weak, so that it is a dwarf star between early-G and late-M. It is
definitely not an L- or T-type object. If cc~2 would be a real companion,
given the magnitude difference between primary and cc~2, the object should be
below the sub-stellar limit with an early-L spectral type (at the same age and
distance as the primary), which we can exclude from our spectrum. Hence, cc~2
is a background object.

\begin{figure}
\resizebox{\hsize}{!}
{\includegraphics[angle=270,width=0.5\textwidth]{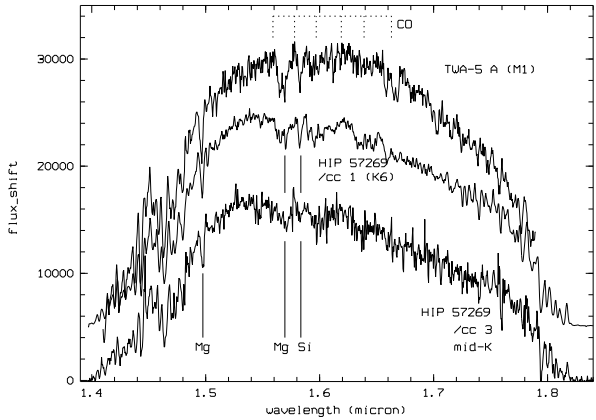}}
\caption{Spectra taken with ISAAC at the VLT in March 19th 2001. For the
  alignment of the slit see Fig.~\ref{fig4}. We show a spectrum of TWA-5~A
  (M1) for comparison and we determine the spectral type of cc~1 to K6 and
  of cc~3 to mid-K.}
\label{fig:specISAAK}
\end{figure}

Then, we took H-band spectra of cc 1 and 3 with the Infrared Spectrograph and
Array Camera (ISAAC, 1024 by 1024 ESO-Hawaii chip) at the ESO 8.2\,m telescope
Antu, Unit Telescope No. 1 (UT1) of the Very Large Telescope (VLT) on Cerro
Paranal on 19 March 2001 between 07:20\,h and
08:16\,h UT in service mode, 28 spectra with 60 sec each through a 1" slit. The
slit was aligned such that both cc~1 and cc~3 were on the slit, but that the
bright primary \object{HIP~57269~A} was outside of the slit. In the
acquisition image and during the spectroscopy, the seeing was around 0.4" to
0.5". Darks, flats, arcs, and spectrophotometric standards were taken in the
same night. Data reduction was done in the standard way: dark subtraction,
normalization, flat fielding, sky subtraction, wavelength calibration, and
co-adding the spectra. Both objects are clearly detected in the final co-added
spectrum, see Fig.~\ref{fig:specISAAK}.

The companion candidate \object{HIP 57269/cc 1} was shown to be K5-7 in
Fig.~\ref{fig3} above, using the optical DFOSC spectrum. This is consistent
with the IR spectrum, where we can see the typical Mg and Si lines and the CO
bands; and the peak emission is more in the blue than in, e.g., the young M1
dwarf TWA-5~A shown as comparison (taken from Neuh\"auser et
al. \cite{neuhaeuser2000}), so that \object{HIP~57269/cc~1} is a bit hotter
than TWA-5 A. Because \object{HIP 57269/cc 1} does not show Lithium absorption
(in its optical spectrum, Fig.~\ref{fig3}), it is not a
companion. \object{HIP~57269/cc~3} is very similar to \object{HIP~57269/cc~1},
with the Mg and Si being slightly weaker, i.e. of spectral type mid-K. Given
the spectral type (mid-K) and its faintness in the IR, \object{HIP~57269/cc~3}
cannot be a companion to \object{HIP~57269~A}; as a true companion, it would
be a low-mass brown dwarf with spectral type L (similar for cc~4, for which we
did not obtain spectra, yet). Hence, \object{HIP~57269/cc~3} is an unrelated
background K star. HP~57269 is a hierarchical triple with a close pair A \& B
and a wide companion C.

\section{Summary}

Is \object{HIP~57269} a member of TWA?  The distance (48.49$\pm$6.54\,pc), the
location in the sky and the spectral type, the position in the H-R diagram,
the $v \sin i$~and X-ray emission are very similar to the other confirmed
TWA-members. It is a visual binary with a known radial velocity companion, in
total a triple system, which is typical for young stars. But the space motion,
as well as the Lithium absorption suggest that it is more likely a young star
belonging to the Pleiades super cluster.  It is clearly a pre-main sequence
star. The other companion candidates cc~1 - cc~4 are likely not members of the
HIP~57269 system nor of TWA.

\acknowledgements

This research has made use of the SIMBAD database, operated at CDS,
Strasbourg, France. R.N. wishes to acknowledge financial support from the
Bundesministerium f\"ur Bildung und Forschung through the Deutsches Zentrum
f\"ur Luft- und Raumfahrt e.V. (DLR) under grant number 50 OR 0003. We would
like to thank the ESO User Support Group for assistance, the ISAAC team for
the VLT service mode observations and also the NTT team with O. Hainaut,
L. Vanci, and M. Billeres for support during the SofI observations.


\end{document}